\documentclass[a4paper,12pt]{article}
\usepackage{amsmath}
\usepackage{graphicx}
\usepackage{rotating}
\usepackage{wasysym}
\DeclareGraphicsExtensions{.pdf,.png,.jpg}
\usepackage{chicago}
\usepackage{klett}
\usepackage{def_the}

\usepackage{Fig_size_A}
\usepackage{kfig}

\renewcommand{\theequation}{\mbox{\arabic{section}.\arabic{equation}}}
\renewcommand{\thefigure}{\arabic{section}.\arabic{figure}}
\renewcommand{\thetable}{\arabic{section}.\arabic{table}}
\renewcommand{\footnoterule}{\rule{14.8cm}{0.3mm}\vspace{+1.0mm}}
\renewcommand{\baselinestretch}{1.0}
\newtheorem{corollary}[theorem]{Corollary}

\title{}
\author{Eckhard Platen}
\begin{document}
\thispagestyle{empty} \vspace*{1.0cm}

\begin{center}
{\LARGE\bf Information-Minimizing Stationary \\\vspace{0.5cm} Financial Market Dynamics } 
\end{center}

\vspace*{.5cm}
\begin{center}

{\large \renewcommand{\thefootnote}{\arabic{footnote}} {\bf Eckhard
Platen}\footnote{University of Technology Sydney,
  School of Mathematical and Physical Sciences, and \\Finance Discipline Group}$^{,}$
}
\vspace*{2.5cm}

\today

\end{center}

\begin{minipage}[t]{13cm} The paper derives the dynamics of a financial market from basic mathematical principles. It models the market dynamics using independent stationary scalar diffusions, assumes the existence of its growth optimal portfolio (GOP), interprets the market as a communication system, and minimizes, in an information-theoretical sense, the joint information of the risk-neutral pricing measure with respect to the real-world probability measure. In this information-minimizing market, its basic independent securities, their sums, minimum variance portfolio, and GOP, as well as the GOP of the entire market, represent squared radial Ornstein-Uhlenbeck processes with additivity and self-similarity properties.

\end{minipage}
\vspace*{0.5cm}


\vspace*{0.5cm}
\vspace*{0.5cm} 

\vspace*{0.5cm}
\noindent{\em Acknowledgements:\/} The author would like to express his gratitude for
\noindent receiving valuable suggestions on the  paper by Mark Craddock, Kevin Fergusson, Len Garces, Martino Grasselli, Matheus Grasselli, Juri Hinz, Hardy Hulley, Sebastien Lleo, Erik Schloegl, Michael Schmutz, Martin Schweizer,   and Stefan Tappe.

	\newpage
	\section{Introduction}\label{section.intro}
 Unifying mathematical principles for the derivation of the laws that determine the  dynamics  of the financial market are  missing in the literature. This paper aims to formulate   such principles and to derive the resulting market dynamics.\\Mathematical principles have been established that facilitate the systematic derivation of the inherent dynamics present in a range of complex natural systems. The  Noether Theorems,  derived by Emmy Noether in \citeN{Noether18}, provide  the fundamental understanding for establishing the respective mathematical principles.\\ 
 The complex dynamical system requires description by partial differential equations (PDEs).   Essential is the identification and optimization of a suitable Lagrangian that captures the impact of the key driver on its dynamics. The presence of Lie-group symmetries for  the solutions of the respective   PDEs   leads to  respective	conservation laws; see  \citeN{KosmannSc18} and \citeN{Olver93}.\\
  	As pointed out in \citeN{KosmannSc18},  the application of the Noether Theorems  requires simplifying assumptions that may  not be perfectly true  but permit the derivation of  laws that permit the successful  engineering of solutions to  practical problems. These laws must be based only on the formulated mathematical principles.   The mathematical principles and their consequences must align with empirical evidence and adhere to logical reasoning.  \\
 To characterize the financial market as a complex stochastic dynamical system via PDEs, the current paper models  a continuous market with normalized independent basic securities, the {\em normalized atoms}, which follow independent stationary scalar diffusions. 
  The normalized atoms evolve  in respective {\em activity times}. Their volatilities  represent stationary processes, which is consistent with empirical evidence; see, e.g., \citeN{Engle82}. Each normalized atom is driven by a unique independent  Brownian motion process. An {\em  atom} is  an auxiliary security  formed by multiplying its normalized value by an exponential function of time.      Therefore, an atom has the same  volatility as the respective normalized atom. The volatility of a normalized atom is defined as a flexible function of its value, multiplied by the square root of its stationary {\em  activity}, which is the derivative of its activity time.  \\ 
	
	The atoms form the stochastic basis and represent the independently fluctuating primary security accounts of the market.    The market of atoms is extended by the  savings account, which has a stationary interest rate, as its locally risk free primary security account. It is assumed to permit continuous trading,   instantaneous investing and borrowing,   short sales with full use of proceeds,    infinitely divisible securities, and      no transaction costs.   Each Brownian motion models a specific economic randomness and is assumed to be indivisible.   The above-described extended market of atoms is called a {\em stationary market} when the  activities, the interest rate, and the growth rate of its {\em minimum variance portfolio} (MVP) are stationary processes.\\
	 The  {\em growth optimal portfolio} (GOP) of the introduced stationary  market  is   interchangeably  called the Kelly portfolio, expected logarithmic utility-maximizing portfolio, or num\'eraire portfolio; 
	 see, e.g.,  \citeN{Kelly56}, \citeN{Merton71}, \citeN{Long90}, \citeN{Becherer01}, \citeN{Platen06ba}, \citeN{KaratzasKa07}, \citeN{HulleySc10}, and \citeN{MacLeanThZi11}.  The existence of the GOP can be interpreted as a  {\em no-arbitrage condition}   because \citeN{KaratzasKa07} 
	 have shown that  the existence of the  GOP is equivalent to their {\em No Unbounded Profit with Bounded Risk} (NUPBR)  condition. This no-arbitrage condition is weaker than the {\em No Free Lunch with Vanishing Risk} (NFLVR)  condition of \citeN{DelbaenSc98}. The request about the existence of the   GOP   is an extremely weak assumption. When violated, the candidate for the GOP would reach infinite values in finite time, which is not the property of a market the current paper aims to model.
	  This leads  to the following first mathematical principle:\\
	 
	 \noindent {\em \bf First Principle}:\\
	{\em For a stationary market, a
		 unique strong solution of the respective system of SDEs, characterizing the market, and  the market's GOP  are assumed to exist.}\\
	 


	     The current paper interprets a financial market as a {\em communication system} in the sense of \citeN{Shannon48}. 
	      The prevailing short-term pricing rule employs the savings account as the num\'eraire.   The {\em joint information},  as defined in  \citeN{Kullback59}, of the  respective risk-neutral pricing measure with respect to the real-world probability measure is defined as the sum of the  self-information of the joint probability density of the normalized atoms and the Kullback-Leibler divergence of the   risk-neutral pricing measure from the real-world probability measure. When this joint information is minimal, all price-relevant details are already reflected in the market dynamics, making its movements most unpredictable and reducing the average squared market prices of risk.   This motivates the formulation of the following second mathematical principle: \\
	      
	     \noindent {\em \bf Second Principle}:\\
	     {\em The stationary market minimizes the joint information of the risk-neutral pricing measure   with respect to the  real-world probability measure.}\\
	      
	       We call a stationary market  an {\em  information-minimizing} market when both principles  apply.     
	    	 It will be shown for an  information-minimizing market that the   information-minimizing  dynamics of atoms, sums of atoms, the GOP of the atoms, the GOP of the entire market, and the MVP are those of squared radial Ornstein-Uhlenbeck (SROU) processes that evolve in respective activity times. SROU processes  are generalizations of the  Cox-Ingersoll-Ross (CIR) process; see   \citeN{CoxInRo85},  \citeN{RevuzYo99}, and \citeN{GoingYo03}.  These  processes  exhibit self-similarity properties in the sense of \citeN{Mandelbrot97}. The information-minimizing market has equal activities and equal average squared market prices of risk, which are determined by the average interest rate and the average growth rate of the MVP in savings account denomination.\\     
	    	 
	    	A large part of the  literature  bases financial market modeling on  expected utility maximization, as described, e.g., in \citeN{Cochrane01} and references therein. The current paper  proposes an alternative way for deriving  realistic market dynamics.   Furthermore, it demonstrates that the  NFLVR condition can be replaced by the weaker condition of the First Principle.\\ 
	    	 Several intuitively appealing  notions of  market efficiency, including those discussed in \citeN{Fama70} and \citeN{GrossmanSt80}, have been studied   in the literature, which do not use the information-theoretical concepts of self-information, joint information, and Kullback-Leibler divergence as employed by the current paper.
	    	 The  information-minimizing market turns out to have realistic properties and  the most unpredictable  market dynamics with the minimal possible average squared market prices of risk. \\ 
    Information minimization is equivalent to entropy maximization; see \citeN{Shannon48}.
     The  mathematical principle of   entropy maximization 
   has played a key role in uncovering  laws of nature across various fields; see  \citeN{KosmannSc18}.  
       By minimizing  the information as a Lagrangian, the normalized atom dynamics become specified. The resulting optimal transition probability density is the solution of  a system of PDEs that has  Lie-group symmetries, as described, e.g., in \citeN{Olver93}, \citeN{CraddockPl04}, and  Chapter 4 in \citeN{BaldeauxPl13}.  The Noether Theorems predict that these Lie-group symmetries  determine the information-minimizing dynamics,  special market properties, and conservation laws.\\
	 	 	 A real market's dynamics generally approximate, but do not exactly match, those of an information-minimizing market.
	 	 	    One can  generalize  the information-minimizing market model by making its parameters  flexible and time-dependent. The resulting market model is likely to be realistic and retain several features of the information-minimizing market model.   This model will be examined in future work.\\ 

	     The paper is organized as follows:  Section 2  introduces  the  financial market. Section 3 minimizes   information and reveals the  information-minimizing market dynamics  and market properties. 
	      Several appendices prove the obtained results.

	        \section{
	       Financial	Market }\setcA \setcB \label{Section2}
	      This section models  a continuous financial market with stationary scalar diffusions modeling the normalized atoms. 
	          \subsection{Atoms}
	        
	          The modeling is performed on a filtered probability space $(\Omega,\mathcal{F},\underline{\cal{F}},P)$, satisfying the usual conditions; see, e.g., 
	          \citeN{KaratzasSh98}.  The filtration $\underline{\cal{F}}$ $=(\mathcal{F}_t)_{t \in [0,\infty)}$ models the evolution of all events covered by the  market model.  
	          The events that evolve until time $t \in [0,\infty)$ are encapsulated by the sigma-algebra $\mathcal{F}_t$, which is, in general,  a superset of the sigma-algebra generated by the randomness of the  driving Brownian motions until  time $t$ and the initial values of securities. \\ 
	            
	                The {\em  atoms} 
	        represent  $n\in\{1,2,...\}$ independent, nonnegative auxiliary securities with values denoted by $A^1_t,...,A^n_t$ at time $t\in[0,\infty)$.  The $k$-th atom, $k\in\{1,...,n\}$,   is only driven by the $k$-th Brownian motion $W^k_t$ and reinvests all dividends or other payments and expenses.  The  $n$  independent driving Brownian motions $W^1_t,...,W^n_t$  evolve in  calendar time $t\in[0,\infty)$ under the real-world probability measure $P$ and the filtration $\underline{\cal{F}}$. These Brownian motions are considered indivisible, meaning that their associated randomness cannot be separated. The values of the atoms  are  denominated in units of a currency. \\    Since we are denominating the securities in units of a currency, the {\em savings account} $A^0_t$ is defined as the exponential
	        \begin{equation}\label {A0}
	        	A^0_t=\exp\left\{\int_{0}^{t}r_sds\right\}
	        \end{equation}
	        for  $t \in [0,\infty)$. Here, $r=\{r_t,t \in [0,\infty)\}$ denotes the continuous, adapted, integrable, stationary {\em interest rate} process.  The savings account is not an atom. Under the First Principle, the  market formed by the atoms has a {\em growth optimal portfolio} (GOP), the {\em atom GOP} $S^*_t$. The market of atoms will have, by construction, no {\em locally risk-free portfolio} (LRP), which is a portfolio with zero volatility. \\
	      
	          	       Theorem 3.1 in \citeN{FilipovicPl09} reveals the general structure of a continuous market that has a GOP and no LRP, which includes the market of atoms. The most striking structural property of such a market is the existence of a unique {\em generalized risk-adjusted return} $\lambda^*_t$, which emerges from the growth rate maximization that identifies  the atom GOP $S^*_t$.   The generalized risk-adjusted return $\lambda^*_t$ is assumed to represent a flexible, continuous, adapted,  integrable, stationary process.  Since the market of atoms does not have an LRP, $\lambda^*_t$ is, in general, different from the interest rate $r_t$. \\
	          	       For each independent driving Brownian motion $W^k_t$, $k\in\{1,...,n\}$, the respective {\em $k$-th atom volatility }   is denoted by $\beta^k_t$ and the respective {\em risk premium factor} by $\omega^k_t$. Both processes are assumed to represent  flexible, continuous, strictly positive,  adapted, square integrable, stationary   processes. \\ Without loss of generality, for  $k\in\{1,...,n\}$, we assume   the   {\em $k$-th  atom process} 
	          $A^k=\{ A^k_t,t\in[0,\infty) \}$  to satisfy  the (It\^o-) stochastic differential equation (SDE) \begin{equation}\label{hatB11j21}
	          \frac{d   A^k_t}{ A^k_t}
	          =\lambda^*_tdt+\beta^k_t(\beta^k_t\omega^k_tdt+dW^k_t) 
	         	          \end{equation}
	         for $t\in[0,\infty)$ with  strictly positive $\mathcal{F}_0$-measurable {\em initial $k$-th atom value} $ A^k_0>0$. 
	          \\
	          
	         The value of an atomic process may approach zero, at which point it is assumed to be instantaneously reflected.   
	           The respective security at a given time starts directly after the last hitting time of zero 
	           and exists until  and including the next hitting time of zero. After that time it becomes another security that replaces the former one. This means, at a given time an atom models a  security that lives between its prior and following  hitting times of zero. \\
	         
	         \subsection{Atom GOP}
	       For matrices and vectors $ \bf{x}$ we 
	       denote by $ {\bf{x}}^\top$ their transpose and write $|{\bf{x}}|=\sqrt{{\bf{x}}^\top{\bf{x}}}$.  Moreover, $\b1=(1, \dots, 1)^\top$ is a vector, and we write $\bf 0$ for a zero matrix or vector, where the dimensions follow from the context. Let us denote by $\beta_t$ the diagonal matrix with the atom volatilities at its diagonal and zeros at all off-diagonal elements.  Furthermore, we denote by ${\bf \omega}=(\omega^1,...,\omega^n)^\top$ the vector of constant risk premium factors. This allows us to write the SDE for the vector of  atoms ${\bf{A}}_t=(A^1_t,...,A^n_t)^\top$ in the form \begin{equation} \label{e.2.14}
	        	\frac{d{\bf{A}}_t}{{\bf{A}}_t}=\lambda^*_t{\bf{1}} dt +\beta_t(\beta_t\omega dt+
	         d{\bf{W}}_t)
	        \end{equation}
	        with the vector of strictly positive {\em initial values}  ${\bf{A}}_0$ and  the vector process ${\bf{W}}=\{ {\bf{W}}_t =(W^1_t,\dots,W^{n}_t)^\top,t \in [0,\infty)\}$ of the $n$   independent  driving  Brownian motions. 
	         Here we write $\frac{d{\bf{A}}_t}{{\bf{A}}_t}$ for the $n$-vector of stochastic differentials $(\frac{dA^1_t}{A^1_t},...,\frac{dA^{n}_t}{A^{n}_t})^\top$.
	         By application of the It\^o formula one obtains for a portfolio $S^{\bar\pi}_t$ of atoms with SDE
	         \begin{equation} \label{e.2.10}
	         	\frac{dS^{\bar\pi}_t}{S^{\bar\pi}_t}=\bar \pi_t^\top\frac{d{\bf{A}}_t}{{\bf{A}}_t}
	         \end{equation}
	        and  weight vector   ${ \bar \pi^{}_t}=(\bar\pi^{1}_t,..., \bar\pi^{n}_t)^\top$ the {\em  growth rate}
	         \begin{equation}
	         	g^{\bar\pi}_t=\lambda^*_t+({\bar\pi}^{}_t)^\top \beta_t\beta_t(\omega
	         	-\frac{1}{2}\bar\pi^{}_t)
	         	\end{equation}
	      as the drift of its logarithm  for $t \in [0,\infty)$. The atom GOP $S^*_t$ is the portfolio that maximizes this growth rate and is defined as follows:
	         \begin{definition}\label{GOP}
	         	The atom GOP  $S^{*}_t$   is the positive  portfolio of atoms with maximum growth rate $g^{\bar\pi}_t$ and initial value $S^{*}_0>0$, where its  weight vector   ${  \bar\pi^{*}_t}=(\bar\pi^{*,1}_t,..., \bar\pi^{*,n}_t)^\top$
	         	is a solution of the well-posed $n$-dimensional constrained quadratic maximization problem
	         	\begin{equation}\label{quop3}
	         	\max \left\{g^{\bar\pi}_t 
	         	|\bar\pi^{}_t \in {\bf R}^{n}, ( \bar\pi^{}_t)^\top  {\bf 1}=1\right\},
	         	\end{equation} 
	         	for all $t \in [0,\infty)$.
	         \end{definition}
	      Appendix A derives the following properties for the atom GOP:   
	         \begin{theorem}\label{stockGOP}
	         	For a market of atoms and $t\in[0,\infty)$, 
	          the sum of the risk premium factors is conserved and equals the constant $1$, that is,
	         	\begin{equation}\label{Nt=1}
	         \omega^\top {\bf{1}}	
	         =1.\end{equation}   When constructing the atom GOP, the vector of weights assigned to the atoms corresponds to the vector of risk premium factors 
	         	\begin{equation}\label{pi1}
	         	\bar\pi^{*}_t=\omega,
	         	\end{equation} 
	         	 and the atom GOP satisfies the SDE  
	         	\begin{equation} \label{e.4.122a}
	         	\frac{dS^{*}_t}{S^{*}_t}= \lambda^*_t dt +(\beta_t\omega)^\top(\beta_t\omega dt+d{\bf {W}}_t)
	         	\end{equation}
	           for	$t \in [0,\infty)$  with $S^*_0>0$. 
	         \end{theorem}
	      Via equation \eqref{Nt=1}, the above theorem reveals the important property   that the sum of the risk premium factors is a conserved quantity. We have introduced above a  parametrization of the market dynamics that represents an alternative to the geometric Brownian motion-type parametrization with constant volatilities and constant expected risk premia  popular in the literature. The reason for the chosen parametrization is the following: When denoting in a componentwise analysis for $k\in\{1,...,n\}$ by $\mu^{\bar \pi,k}_t={\bar \pi}^k_t(\lambda^*_t+(\beta^k_t)^2\omega^k)$ the $k$-th component of the expected return of a portfolio $S^{\bar \pi}_t$ of atoms and by $\sigma^{\bar \pi,k}_t={\bar \pi}^k_t\beta^k_t$ the $k$-th component of its volatility, then it emerges from \eqref{e.2.10} for all portfolios of atoms that the ratio 
	       \begin{equation}
	       	\frac{\mu^{\bar \pi,k}_t-{\bar \pi}^k_t\lambda^*_t}{\beta^k_t\sigma^{\bar \pi,k}_t}=\omega^k
	       \end{equation}
	equals   the $k$-th risk premium factor. This means that the risk premium factors are central invariants  of the market of atoms. Even when market conditions and randomness are changing over time, it is not easy to find quantifiable reasons why the risk premium factors would change  over time. Accordingly, this insight and their pivotal role have been acknowledged by the aforementioned parametrization. We introduced a parametrization of the discussed market of atoms, which remains general due to the continued flexibility of the activities involved. The introduced parametrization  will be necessary to allow the full minimization of the joint information requested by the Second Principle. 

	             To prepare later results for a market of atoms, 
	            we  introduce its  {\em minimum variance portfolio} (MVP) $ S^{{MVP}}_{t}$ as follows:
	            \begin{definition}\label{MVP}
	            	The MVP of a market of  atoms  is the positive  portfolio $ S^{{MVP}}_{t}$  with weight vector $\bar \pi^{MVP}_t=(\bar\pi^{MVP,1}_t,...,\bar\pi^{MVP,n}_t)^\top$ and minimum squared volatility \begin{equation}(\sigma^{MVP}_t)^2=\sum_{k=1}^{n}( \bar\pi^{MVP,k}_t \beta^k_t)^2\leq (\sigma^{\bar \pi}_t)^2\end{equation} among all portfolios $S^{\bar\pi}_t$ of atoms and initial value  $ S^{{MVP}}_0=\sum_{k=1}^{n}A^k_{0}$, where its weight vector   ${ \bar\pi^{MVP}_t}$
	            	is a solution of the well-posed $n$-dimensional constrained quadratic minimization problem
	            	\begin{equation}\label{quop2}
	            		\min \left\{(\sigma^{MVP}_t)^2| \bar\pi^{MVP}_t \in {\bf R}^{n}, (\bar\pi^{MVP}_t)^\top  {\bf 1}=1\right\},
	            	\end{equation} 
	            	for all $t \in [0,\infty)$.
	            \end{definition}
	           
	           \subsection{Stationary Volatilities}
	  The Noether Theorems require the characterization of the  financial market dynamics  via a set of PDEs. By following the First Principle, we achieve this by modeling the normalized atoms as independent scalar diffusion processes that evolve in respective  activity times. The probability densities of the normalized atoms are characterized by  respective Fokker-Planck equations or Kolmogoroff PDEs; see \citeN{KaratzasSh98}. We assume for  $k\in\{1,...,n\}$ the {\em  $k$-th normalized  atom} $Y^k_{\tau^k_t}$ to be given in the form
	          \begin{equation}\label{Yktayt1}
	          Y^k_{\tau^k_t}=\frac{A^k_t}{B_te^{\tau^k_t-\tau^k_0}
	          }
	          \end{equation} with {\em basis exponential}
	          \begin{equation}\label{basis}
	         	B_t=\exp\left\{\int_{0}^{t}\lambda^*_sds\right\},
	          \end{equation}
	         and to evolve for $t\in[0,\infty)$ as a  scalar diffusion process in the   {\em $k$-th activity time}
	          \begin{equation}\label{bartaukt}
	          \tau^k_t=\tau^k_0+\int_{0}^{t}a^k_sds,
	          \end{equation}  with strictly positive, continuous, adapted, square integrable {\em $k$-th activity} process $a^k=\{a^k_t,t\in[0,\infty)\}$,   $\mathcal{F}_0$-measurable {\em $k$-th initial activity time} $\tau^k_0$, and {\em $k$-th initial atom value} $A^k_0=Y^k_{\tau^k_0}>0$.  By application of the It\^o formula it follows that a normalized atom has the same volatility as the  respective atom.\\ 

	          For  
	         	          the stationary $k$-th normalized atom process $Y^k=\{Y^k_{\tau^k}, \tau^k\in[\tau^k_0,\infty)\}$, $k\in\{1,...,n\}$, we denote 
	         	           by $p^k_t$ its probability density at the time $t\in[0,\infty)$, where  ${\bf{E}}^{p^k_t}\left(.\right)$ denotes the expectation taken with respect to  this density.  
	         For $k\in\{1,...,n\}$ and $t\in[0,\infty)$, we parametrize 
	          the {\em $k$-th arithmetic mean}
	          \begin{equation}\label{barY0*}
	         {\bf{E}}^{p^k_t}\left(Y^k_{\tau^k_t}\right)=\bar Y^k>0
	          \end{equation}
	          and  
	          the {\em $k$-th logarithmic mean}  
	          \begin{equation} \label{barstalogmean*}
	          	{\bf{E}}^{p^k_t}(\ln(Y^k_{\tau^k_t}))=\zeta^k\in(-\infty,\infty), \end{equation}
	          where we assume $\bar Y^k$ and $\zeta^k$ to represent  flexible constants. \\

	       
	       Without loss of generality, we obtain for $k\in\{1,...,n\}$   general stationary scalar diffusion dynamics of the $k$-th normalized atom $Y^k_{\tau^k_t}$ in the $k$-th activity time $\tau^k_t$ by modeling its volatility  in the form
	          \begin{equation}\label{beta}
	          \beta^k_t=\sqrt{\frac{a^k_t}{\phi^k(Y^k_{\tau^k_t})}}
	          \end{equation} 
	      for $t\in[0,\infty)$.    The {\em $k$-th volatility function}  $\phi^k(.)$ is assumed to be  a   flexible, infinitely often continuously differentiable, strictly positive function of the value of the $k$-th normalized atom  such that a unique strong solution of the resulting SDE for the $k$-th normalized atom in the $k$-th activity time exists; see, e.g., Section 7.7 in \citeN{PlatenHe06}. By  \eqref{Yktayt1}, \eqref{hatB11j21}, \eqref{bartaukt}, and application of the It\^o formula the $k$-th normalized atom $Y^k_{\tau^k_t}$ satisfies  the SDE 
	          \begin{equation}\label{bardYtau}
	          d Y^k_{\tau^k_t}=Y^k_{\tau^k_t}\left(\phi^k(Y^k_{\tau^k_t})^{-1}\omega^k-1\right)a^k_t dt+Y^k_{\tau^k_t}\phi^k(Y^k_{\tau^k_t})^{-\frac{1}{2}}\sqrt{a^k_t}d W^k_{t}
	          \end{equation}
	         with instantaneous reflection at zero for $t\in[0,\infty)$. 
	          There exists an ambiguity in Equation \eqref{beta} when specifying the volatility $\beta^k_t$  by the activity $a^k_t$ and the properties of $Y^k_{\tau^k_t}$. We exploit this ambiguity and remove it by fixing  the arithmetic mean of $Y^k_{.}$ as
	          \begin{equation}\label{omegakk}
	          	\bar Y^k=\omega^k
	          \end{equation}
	         for $k\in\{1,...,n\}$ and $t\in[0,\infty)$. Since   the activity is a flexible, continuous, positive, adapted, square integrable stationary process,  the above-introduced volatility process  can  model  any strictly positive, adapted, square integrable stationary volatility process.  
	            \subsection{Stationary Market }
	        We obtain the
	          {\em stationary market } by adding to the $n$ atoms $A^1_t,...,A^n_t$ as 
	          primary security accounts
	          the  savings account $A^{0}_t$ as a primary security account.  
	           For a portfolio $S^{\pi}_t$ of the primary security accounts $A^0_t,...,A^n_t$ in the stationary market with   weight vector   ${  \pi^{}_t}=(\pi^{0}_t,..., \pi^{n}_t)^\top$ and SDE
	          \begin{equation} \label{e.2.15}
	          	\frac{dS^{\pi}_t}{S^{\pi}_t}=\sum_{k=0}^{n}\pi^k_t\frac{dA^k_t}{A^k_t},
	          \end{equation}
	         one obtains, similarly as for the market of atoms,   its growth rate
	         \begin{equation}
	         	g^{\pi}_t=\lambda^*_t(1-\pi^0_t) +r_t \pi^0_t+\pi_t^\top \beta_t\beta_t(\omega 
	         	-\frac{1}{2}\pi^{}_t)
	         \end{equation}  
	          as the drift of the SDE for the logarithm  of this portfolio at time $t \in [0,\infty)$. We denote by \begin{equation}
	          	G^{S^{\pi}}_t={\bf{E}}^{p_t}(g^\pi_t)
	          	\end{equation}  its {\em average growth rate} at time $t\in[0,\infty)$  and by 
	          	\begin{equation}
	          		G^{S^{\pi}}={\bf{E}}(G^{S^{\pi}}_t)
	          	\end{equation} its {\em expected growth rate}.  In analogy to Definition \ref{GOP}, the following notion is introduced:
	         \begin{definition}\label{GOP1}
	        	For a stationary market, the {\em extended market GOP}  $S^{**}_t$   is the positive  portfolio of savings account and atoms with maximum growth rate $g^{\pi^{**}}_t$ and initial value $S^{**}_0>0$, where its  weight vector   ${  \pi^{**}_t}=(\pi^{**,0}_t,\pi^{**,1}_t..., \pi^{**,n}_t)^\top$
	        	is a solution of the well-posed $(n+1)$-dimensional constrained quadratic maximization problem
	        	\begin{equation}\label{quop3}
	        		\max \left\{g^{\pi}_t 
	        		|\pi^{}_t \in {\bf R}^{n+1}, \pi_t^\top  {\bf 1}=1\right\},
	        	\end{equation} 
	        	for all $t \in [0,\infty)$.
	        \end{definition} The  First Principle  requires the existence of the extended market GOP. Therefore, the following result is derived in Appendix B:
	          \begin{theorem}\label{GOPentire}
	          For a 
	          stationary market, the  extended market GOP $S^{**}_t$ satisfies the SDE
	          \begin{equation} \label{bare.4.113}
	          \frac{d S^{**}_t}{ S^{**}_t}=r_t dt +
	          \theta_t^\top (\theta^{}_tdt+d{\bf W}_t)
	          \end{equation}
	          with initial value $ S^{**}_0>0$,  {\em  market price of risk vector}
	          \begin{equation}\label{thetak}
	          	\theta^{}_t=(\lambda^*_t-r_t) \beta_t^{-1}{\bf{1}}+\beta_t\omega, 
	          \end{equation}
	          extended market GOP-weight vector $\pi^{**}_t=(\pi^{**,0}_t,  \pi^{**,1}_t,..., \pi^{**,n}_t)^\top$ with weights
	          \begin{equation}
	          	\bar\pi^{**}_t=( \pi^{**,1}_t,..., \pi^{**,n}_t)^\top=(\lambda^*_t-r_t) \beta_t^{-2}{\bf{1}}+\omega,
	          \end{equation} to be invested in the atoms $A^1_t,...,A^n_t$, and the weight
	          \begin{equation}\label{rt3}
	          	\pi^{**,0}_t
	          	=(r_t-\lambda^*_t){\bf{1}}^\top\beta_t^{-2}{\bf{1}} 
	          \end{equation}
	          to be invested in the savings account $A^0_t$ 
	          for  	$t\in[0,\infty)$.
	          \end{theorem}
	          Since the activity processes 
	          remain flexible and  Theorem 3.1 in \citeN{FilipovicPl09}  provides necessary and sufficient conditions for the structure of a continuous market,  it follows  that there exists an extremely wide range of continuous extended market dynamics with the same market price of risk processes and extended market GOP that can be transformed into the introduced stationary market model by forming the atoms as respective portfolios of stocks and savings account. 
	          
	          \section{Information-Minimizing Market}\setcA \setcB
	          In this section,  we  interpret an extended market as a {\em communication system} and minimize the joint information of  the risk-neutral short-term pricing measure with respect to the real-world probability measure.  
	          \subsection{Information-Minimizing Market Theorem}

	           	  {\em Information} here is measured as in information theory, following  \citeN{Shannon48} and \citeN{Kullback59}: 
	           	  \begin{definition}
	           	   	A continuous density $q$ of a real-valued  random variable  provides, with respect to a continuous probability density $p$ of a real-valued random variable, the {\em information} 
	           	   \begin{equation}\label{Ipq}
	           	   	\mathcal{I}(p,	q)=	
	           	   	\int_{0}^{\infty} p(y)\ln(q(y))dy,
	           	   \end{equation} 
	           	where $\mathcal{I}(p,	p)$ is called the {\em self-information} of $p$,   and one has the {\em joint information} 
	           	   \begin{equation}\label{jointinfo}
	           	   	\mathcal{I}(p,	\Lambda)=	\mathcal{I}(p,	p)+\mathcal{I}(p,	q)
	           	   		           	   \end{equation} of $q$ with respect to $p$, where	 the {\em Radon-Nikodym derivative} of $q$ with respect to $p$ is  denoted by $\Lambda$.
	           	   	           	   The   {\em  Kullback-Leibler divergence }
	           	   of a time-dependent density $q_t$ with respect to a time-dependent probability density $p_t$ is defined as 
	           	   \begin{equation}
	           	   	I(p_t,	q_t)=	\frac{d}{dt}\mathcal{I}(p_t,	q_t)\end{equation}
	           	   for $t\in[0,\infty)$, as long as the above quantities exist.
	           	\end{definition}
	           	 A random variable is most unpredictable when the information of its density is minimized for the given parameterization; see \citeN{Shannon48}.  Intuitively, the market participants are pricing into the traded prices all available information. When information is minimized, no extra data remains for potential benefit.
	           	  \\ It is reasonable to assume that short-term risk-neutral pricing   is performed when trading, which employs the savings account $A^0_t$ as the num\'eraire and the {\em risk-neutral pricing measure} $Q$ as the pricing measure; see, e.g.,  \citeN{Jarrow22}. 	The latter is characterized by the
	           	   {\em risk-neutral  Radon-Nikodym derivative} 
	           	   \begin{equation*}
	           	   	\Lambda_t=\frac{dQ}{dP}\Big |_{\mathcal{F}_t}=\frac{d(\frac{  A^0_t}{S^{**}_t})}{\frac{  A^0_0}{S^{**}_0}}
	           	   \end{equation*}
	           	  for $t\in[0,\infty)$; see, e.g., Section 9.4 in \citeN{PlatenHe06}.
	           	  For the given  stationary market, the joint information of $Q$ with respect to $P$ at time $t\in[0,\infty)$ amounts to  
	           	  $
	           	  \mathcal{I}(p_t,	\Lambda_t)$. \\
	           	   Trades are discrete, and the   trading intensities, which quantify the intensities of the flow of  price information,  are modeled by the respective activities in the above-introduced stationary extended market model.  A trade at time $t\in[0,\infty)$ reveals information through its price. The risk-neutral joint  density  of the  normalized atoms in their respective activity times at time $t\in[0,\infty)$ is denoted by $q_t=\prod_{k=1}^{n}q^k_{t}$, and    the respective real-world joint probability density    is given by $p_t=\prod_{k=1}^{n}p^k_{t}$.    
	           	    We denote by $p^k_{\infty}$ the stationary density of the $k$-th normalized atom  and by $p_{\infty}=\prod_{k=1}^{n}p^k_{\infty}$  their  stationary joint density. \\
	           	   
	           	    It should be emphasized that the available information about the price of a normalized atom is already fully priced in when its self-information is minimized. This means,  \textquoteleft no surprises' will occur when the normalized atom follows the respective  stationary dynamics. By Equation \eqref{jointinfo},  the {\em joint information of the risk-neutral density $q_t$ with respect to $p_t$} at time $t\in[0,\infty)$ we define as
	           	   \begin{equation}
	           	   		\mathcal{I}(p_{t},	\Lambda_t)=	\mathcal{I}(p_{t},	p_{t})+\mathcal{I}(p_{t},	q_{t})=	\mathcal{I}(p_{0},	p_{0})+\int_{0}^{t}I(p_{s},	q_{s})ds.
	           	   \end{equation}
	           	 To prepare its minimization, as requested by the Second Principle, we introduce the {\em average activity} 
	           	 \begin{equation}\label{averactivity}
	           	 	a_t=\left(\sum_{k=1}^{n}\omega^k\sqrt{\frac{1}{a^k_t}}\right)^{-2}
	           	 \end{equation} at time $t\in[0,\infty)$. In reality, the average activity  is  much more rapidly moving than the normalized atoms. A detailed model of its dynamics will be presented in future work.  \\  We introduce the following notion:
	           	   \begin{definition} A stationary   market 
	           	   	 is said to be an {\em information-minimizing market} 
	           	   if	the joint information  $
	           	   	\mathcal{I}(p_{t},	\Lambda_t)$ of its risk-neutral density $q_t$ with respect to the real-world probability density $p_t$ is minimized for all $t\in[0,\infty)$.
	           	   \end{definition} 
	           	 An information-minimizing market offers a clear, mathematically defined concept of market efficiency, distinct from those in \citeN{Fama70} and \citeN{GrossmanSt80}.     The main difference from these notions arises from the fact that the above definition does not focus on any moments of prices. Instead, it takes the entire probability density with information-theoretical quantifications into account, which yields realistic market dynamics after minimization. 
	           \\
	           	
	           	 Appendix C derives the following theorem that summarizes how  the minimization of the joint information of  $q_t$ with respect to $p_t$ determines the market dynamics:
	           	   		\begin{theorem}\label{informationtheorem} For $k\in\{1,...,n\}$ and $t\in[0,\infty)$, the dynamics of the $k$-th normalized atom $Y^k_{\tau^k_t}$ of an information-minimizing  market is that of a {\em square root process  of dimension $\frac{4}{n}$  that evolves  in the $\tau^k$-time}
	           	   			\begin{equation}
	           	   				\tau^k_t=\tau^k_0+\hat \tau_t
	           	   			\end{equation}  with {\em average  activity time} 
	           	   			\begin{equation}
	           	   			\hat	\tau_t=\int_{0}^{t}a_sds,
	           	   			\end{equation}
	           	   		equal activities \begin{equation}\label{akt}
	           	   			a^k_t= a_t,\end{equation} 
	           	   			and  equal risk premium factors \begin{equation}\label{omega}\omega^k=\frac{1}{n},\end{equation} 
	           	   		satisfying the SDE
	           	   			\begin{equation*}\label{dY2}
	           	   				dY^k_{\tau^k_t}=\left(\frac{1}{n} -Y^k_{\tau^k_t}\right)a_tdt+ \sqrt{  Y^k_{\tau^k_t}a_t} d W^k_t
	           	   			\end{equation*}	
	           	   			with initial value $
	           	   			Y^k_{\tau^k_0}=A^k_0$,
	           	   			distributed according to the information-minimizing stationary density $ \bar p^{k}_{t}=\bar p_{t}=\bar p_{\infty}$, which is  a gamma density with $\frac{4}{n}$ degrees of freedom and mean $\frac{1}{n}$. The minimized Kullback-Leibler divergence of the risk-neutral density $q_t$ with respect to the  stationary real-world density $\bar p_t$ yields 
	           	   			 the {\em  information-minimizing generalized risk-adjusted return }
	           	   			\begin{equation}\label{hatr}
	           	   			\lambda^*_t= r_t+\hat \lambda a_t  \end{equation} 
	           	   			 with constant {\em net-risk-adjusted return in activity time}
	           	   			 \begin{equation}\label{hatlambda}
	           	   			 \hat \lambda=\frac{G^{S^{MVP}}-G^{A^0}}{ {\bf{E}}(a_t)}-1
	           	   			 	\end{equation} and the {\em minimized Kullback-Leibler divergence}
	           	   			\begin{equation}\label{relinfo}
	           	   		I(\bar p_{t},q_t)=G^{S^{**}}-G^{A^0}=\frac{1}{2}{\bf{E}}\left( a_t\right) \left(\hat \lambda^2+\bar \omega(n) +2\hat \lambda\right)
	           	   	\end{equation}
	           	   			equal to the
	           	   			 average growth rate
	           	   			 of the extended market GOP in savings account denomination, where 
	           	   	           	   		 \begin{equation}\label{baromega}
	           	   				\bar\omega(n)={\bf{E}}^{\bar p^1_t}\left(\frac{ \frac{1}{n} }{Y^1_{\tau^1_t}}\right)
	           	   			\end{equation}
	           	   	denotes the	{\em average squared atom GOP volatility in activity time}. 
	           	   		
	           	   		\end{theorem}
	           	   		By applying the Second Principle, which means when the joint information $\mathcal{	I}( p_{t},\Lambda_t)$ is minimized, this theorem reveals the optimal market dynamics.	It shows that the information-minimizing market is minimizing the self-information of the stationary densities of the normalized atoms and the Kullback-Leibler divergence of the risk-neutral densities  with respect to the real-world densities.   The minimized Kullback-Leibler divergence equals the  expected	growth rate  of the  GOP of the extended market in savings account denomination. The minimized self-information of $p_t$ yields the most unpredictable market dynamics.\\
	   	           	    The generalized risk-adjusted return $\lambda^*_t$ is a Lagrange multiplier. It is shown to be the sum of the prevailing interest rate $r_t$ and the product of the  average activity $a_t$ and the net-generalized risk-adjusted return in activity time  $\hat \lambda$. To minimize fully the Kullback-Leibler divergence in the proof of the above theorem, the  net-generalized risk-adjusted return in activity time must be a constant. This constant  is  a  macro-economic parameter. It is determined by the average productivity of the  economy and its average interest rate.\\ As shown in Theorem 7.1 in \citeN{FilipovicPl09}, the central bank can set the interest rate freely without violating the First Principle. Forthcoming work will show how the setting of the interest rate influences the inflation rate and how the interest rate can be set  to optimally benefit  the consumption.\\
	           	    	 Since the observed values and fluctuations of the atoms   do not provide any \textquoteleft surprises',  the information-minimizing market is, in some sense, \textquoteleft efficient'. However, this kind of market efficiency is different from the notions of market efficiency  discussed in  the literature on \textquoteleft efficient capital markets' or \textquoteleft informationally-efficient' markets; see, e.g., \citeN{Fama70} and \citeN{GrossmanSt80}. The key difference compared to these notions  is  that  the current paper relies fully on an information-theoretical definition of information. It is this choice of the notion of market efficiency that leads  to    realistic market dynamics. 
	           	    Generally, developed markets are not fully information-minimizing, but most likely often come close.  Forthcoming work will derive a conservation law by studying a market that keeps the  information-minimizing form of the volatility functions $\phi^k(y)=y$ for all $k\in\{1,...,n\}$, makes the risk premium factors time-dependent, and models the  atom activities proportional to the average activity with time-dependent weights.  
	           	   	
	         \subsection{Information-Minimizing Atom Dynamics}  	   			
	           	   		
	          For  $k\in\{1,...,n\}$, it follows from Theorem \ref{informationtheorem} that the information-minimizing $k$-th normalized atom process $Y^k_{.}$ evolves in the $k$-th  activity time $\tau^k$
	       as   a square root process, as described, e.g., in Section 4.4 of \citeN{PlatenHe06}. This process, called the Cox-Ingersoll-Ross (CIR) process, gained prominence in finance through \citeN{CoxInRo85}.   It is alternatively known as a squared radial Ornstein-Uhlenbeck (SROU) process,   as described in \citeN{RevuzYo99} and 
	       	\citeN{GoingYo03}, with  dimension
	          $ d_k=4\omega^k$
	          and    arithmetic mean $\omega^k =\frac{1}{n}$.  
	                  	                  Its volatility   is, by  construction, the same volatility as the volatility of the $k$-th atom 
	           	         \begin{equation}\label{Akt2}  A^k_{t}=Y^k_{\tau^k_t}B_t e^{\tau^k_t-\tau^k_0},
	           	          \end{equation}  which has by \eqref{hatB11j21},  \eqref{Yktayt1}, and application of the It\^o formula, the following properties:
	           	           
	           	          \begin{corollary}\label{corolvarphiAk}
	           	          	For  
	           	          	$k\in\{1,...,n\}$, the  $k$-th atom in an information-minimizing market has the dynamics  of  an  SROU  process with {\em dimension} $d_k=\frac{4}{n}$. 
	           	            It evolves in the   $k$-th intrinsic time $\varphi^k(t)$, which has the time derivative 
	           	          	 \begin{equation}\label{varphiAkt}
	           	          	 \frac{d\varphi^k(t)}{dt}=\frac{ B_te^{\int_{0}^{t}a_sds} a_t}{4}
	           	          	 \end{equation} and satifies the SDE
	           	          	         	 \begin{equation}\label{Akt} d   A^k_{t}=(r_t+\hat \lambda a_t )A^k_tdt+
	           	          	 	\frac{4}{n}\frac{d\varphi^k(t)}{dt}dt+\sqrt{ A^k_t}\sqrt{4 \frac{d\varphi^k(t)}{dt}}dW^k_t   
	           	          	 \end{equation}
	           	           for $t \in[0,\infty)$,	 with             	          	  random  initial value $ A^k_{0}$ with density $p^k_0=p_\infty$ 
	           	          	.
	            \end{corollary} 
	          	         
	          Key to the understanding of the nature of the information-minimizing dynamics of an atom is the proportionality of the square of its diffusion coefficient to its value in its SDE \eqref{Akt}. This proportionality arises because capital evolves as the sum of numerous independently changing investment units.  These capital units independently generate new capital units or vanish.  By the fundamental mathematical fact that the variance of the sum of independent random variables equals the sum of their variances, over a short period, the variance of the increment of the sum of  independently evolving capital units becomes proportional to the original number of the capital units at the beginning of the short period. The continuous limit of these dynamics produces a diffusion coefficient in the corresponding SDE for the total capital units, which is proportional to the sum's value, as indicated in SDE \eqref{Akt}.    The above dynamics are  analogous to the limiting dynamics of  population sizes modeled by branching processes, as described  in \citeN{Feller71}. The continuous limits of the dynamics of branching processes are those of SROU processes; see \citeN{Feller71} and \citeN{GoingYo03}. 
	          \\ \citeN{CraddockPl04} examined Lie-group symmetries in systems where the diffusion coefficient is proportional to the square root of the state variable.  By taking the particular SDE \eqref{Akt} of atoms into account, it follows from Theorem 4.4.3 in \citeN{BaldeauxPl13} an explicit formula for the transition probability density of an atom. It is that   of an SROU  process, as shown in the derivation of the Equation (5.1.2) of the  monograph  \citeN{BaldeauxPl13}. Therefore, the transition probability density for the dynamics of a normalized atom results from a Lie-group symmetry, as  expected from the Noether Theorems; see \citeN{Noether18}.\\

          	\subsection{Additivity Property of Sums of Atoms}

          	 The sum of  squared Bessel processes is known to form again a squared Bessel process; see \citeN{ShigaWa73}. This {\em additivity property}  results from the special form of the PDE of the transition probability density of  a squared Bessel process. 
          	          	The following result is obtained by applying Corollary \ref{corolvarphiAk}: 
          	 
          	\begin{corollary}\label{sumofatoms}
          		For  an information-minimizing market 
          		 and a set $\mathcal{A}\subseteq \{1,...,n\}$ of indexes of atoms,  the respective {\em sum of  atoms }
          		\begin{equation}
          		 A^{\mathcal{A}}_{t} =\sum_{k\in\mathcal{A}}  A^k_t         \end{equation} satisfies the SDE
          			\begin{equation}\label{Akphi1}
          		d  A^{\mathcal{A}}_{t}
          		= \lambda^*_tA^{\mathcal{A}}_{t}dt+
          		d_{\mathcal{A}}d\varphi^{\mathcal{A}}(t)+2\sqrt{A^{\mathcal{A}}_{t} \frac{d\varphi^{\mathcal{A}}(t)}{dt}}d W^{\mathcal{A}}_t\end{equation}
          		of an SROU   process in the {\em $\mathcal{A}$-intrinsic  time}
          		\begin{equation}
          			\varphi^{\mathcal{A}}(t)=\varphi^{\mathcal{A}}(0)+\frac{ 1}{4}\int_{0}^{t}  B_se^{\int_{0}^{s}a_zdz}a_sds
          		\end{equation} with generalized risk-adjusted return $\lambda^*_t$, dimension 
          		\begin{equation}
          		d_{\mathcal{A}}=4\sum_{k\in\mathcal{A}}\frac{1}{n},
          		\end{equation}
          and	initial value
          		\begin{equation}
          		 A^{\mathcal{A}}_{0} =\sum_{k\in\mathcal{A}}  A^k_0 ,
          		\end{equation}
          		where $ W^{\mathcal{A}}_t$ is a Brownian motion with stochastic differential
          		\begin{equation}
          		d W^{\mathcal{A}}_t=\frac{1}{\sqrt{  A^{\mathcal{A}}_{t} }}\sum_{k\in\mathcal{A}} \sqrt{ A^k_{ t}}d W^k_t
          		\end{equation}
          		for $t\in[0,\infty)$ and initial value $ W^{\mathcal{A}}_0=0$.
          		The respective {\em normalized sum of atoms} 
          		\begin{equation}\label{YAP}
          			Y^{\mathcal{A}}_{\tau^{\mathcal{A}}_t}=\frac{	S^{\mathcal{A}}_{t}}{B_te^{\int_{0}^{t}a_sds}}=\sum_{k \in {\mathcal{A}}}Y^k_{\tau^k_t}
          		\end{equation}
          		satisfies the SDE
          		\begin{equation}
          			\label{Ykappa5}
          			dY^{\mathcal{A}}_{\tau^{\mathcal{A}}_t}	=\left(\frac{d_{\mathcal{A}}}{4} -Y^{\mathcal{A}}_{\tau^{\mathcal{A}}_t}\right)a_tdt+\sqrt{  Y^{\mathcal{A}}_{\tau^{\mathcal{A}}_t}a_t} d  W^{\mathcal{A}}_t
          		\end{equation}
          		and evolves as a square root process of dimension $d_{\mathcal{A}}$ in the $\mathcal{A}$-activity time 
          		\begin{equation}
          			\tau^{\mathcal{A}}_t=\tau^{\mathcal{A}}_0+\hat \tau_t
          		\end{equation}
          		for $t\in[0,\infty)$.\\
                   		           	\end{corollary}
          	
          	The above additivity property is a fundamental property of  sums of atoms of an information-minimizing market. Each sum of atoms forms  a squared radial Ornstein-Uhlenbeck process with the sum of the dimensions of the summands as its dimension and the respective sum of the initial values as its initial value.
          	          	 One notices that the above result characterizes a self-similarity property in the sense of \citeN{Mandelbrot97}. In this case, the transition probability density of a sum is of the same type as those of its summands.  \\
          	          	 
          A special sum of atoms is the sum of all atoms $S^{AP}_t$, which we call the {\em  atom portfolio} (AP). By application of Corollary \ref{sumofatoms}, the It\^o formula, and Equation \eqref{Nt=1}, one can draw directly the following conclusion:
           	 \begin{corollary}\label{sumatomAP}
          	 	 For  an information-minimizing market, the  atom portfolio
          	\begin{equation}\label{SAP}
          	S^{AP}_{t}=\sum_{k=1}^{n}  A^k_{t}, \end{equation}
          	follows a time-transformed SROU process of dimension four. 
          	 	           	 \end{corollary} 
          	 	           	 	For an information-minimizing market, 
          	 	           	 the sum of atoms  evolves in a respective intrinsic  time as an SROU  process 
          	 	           	 with  squared volatility   proportional to the inverse of
          	 	           	 the  normalized sum of atoms.
          	 	           	 This inverse follows    a stationary square root process or CIR process 
          	 	           	 in its activity time. With respect to this time, the normalized sum of atoms  has        a gamma density as its stationary density, where its dimension $d_{\mathcal{A}}$ equals the degrees of freedom. \\	 
          	 	           	 It is well  known that the Student-$t$ density with degrees of freedom $d_{\mathcal{A}}> 0$  is the  normal-mixture density that emerges as the log-return density when the mixing stationary density of the squared volatility is the inverse of a gamma density with $d_{\mathcal{A}}$ degrees of freedom; see, e.g., \citeN{HurstPl97d} and \citeN{PlatenRe08e}.     	             If one would interpret the market portfolio of the world stock market as its AP 	 and would estimate the log-return density of  the market capitalization-weighted MSCI world stock index, then the above theoretical property would predict the estimation of a Student-$t$ density with  about four degrees of freedom. \citeN{FergussonPl06dc} found that, when tested across a broad spectrum of possible log-return densities, this hypothesis is not readily dismissed.\\   Due to the above-derived additivity property of  atoms, the dimension of a market capitalization-weighted country stock index, when interpreted as a sum of  atoms, may be slightly lower than that of the world stock index because some  atoms that drive the stock indices of other countries would not be included in the country stock index. In several independent studies, including  \citeN{MarkowitzUs96a}, \citeN{MarkowitzUs96b}, and  \citeN{HurstPl97d},  the hypothesis could not be easily rejected that log-returns of market capitalization weighted total return stock  indices of countries have a Student-$t$ density   with about four or slightly less degrees of freedom  when tested in a large class of potential log-return densities.  Furthermore, in \citeN{BreymannLuPl09e} it has been shown that  the hypothesis of approximately four to five degrees of freedom of the Student-t density cannot be easily rejected as the typical log-return density of a  world stock index when denominated in currencies.  The empirical evidence documented in the mentioned papers supports   a  diffusion model of the derived type.  \\  
          	 	           	 	 \subsection{Atom GOP Dynamics}
          	 	           	 The following dynamics of the atom GOP are derived in Appendix D:
          	 	           	 \begin{theorem} \label{stockGOP1}
          	 	           	 	For  an information-minimizing market 
          	 	           	  the  atom GOP $S^*_t$
          	 	           	 	satisfies  the SDE
          	 	           	 	\BE \label{e.4.118}
          	 	           	 	d S^*_{t}= \lambda^*_t S^*_{t}dt+
          	 	           	 	4\frac{d\varphi^*(t)}{dt}dt+2 \sqrt{ S^*_{t}  \frac{d\varphi^*(t)}{dt}} dW^*_t \EE
          	 	           	 	with $ S^*_{0}>0$. It is investing with equal weights in the atoms and evolves as an SROU process in the  {\em intrinsic atom-GOP   time} 
          	 	           	 	\begin{equation}\label{varphi*}
          	 	           	 		\varphi^*({t})
          	 	           	 		=\varphi^*({0})+\frac{1}{4}\int_{0}^{t}S^*_sZ_sa_sds
          	 	           	 	\end{equation}
          	 	           	 	with squared atom-GOP volatility in activity time \begin{equation}\label{Zt}
          	 	           	 		Z_t= \frac{1}{n} \sum_{k=1}^{n}\frac{\frac{1}{n}  }{Y^k_{\tau^k_t}},
          	 	           	 	\end{equation} 
          	 	           	 	where $W^*_t$ is  a Brownian motion  with stochastic differential
          	 	           	 	\begin{equation}\label{W*t}
          	 	           	 		dW^*_t= Z_t^{-\frac{1}{2}}\sum_{k=1}^{n}\frac{\frac{1}{n} }{\sqrt{Y^k_{\tau^k_t}}}d W^k_t
          	 	           	 	\end{equation} and initial value $W^*_0=0$          
          	 	           	 	for $t \in [0,\infty)$.
          	 	           	 	The normalized atom GOP
          	 	           	 	\begin{equation}\label{Y*1}
          	 	           	 		Y^*_{\tau^*_t}=\frac{ S^*_{t}}{B_te^{\tau^*_t-\tau^*_0}}
          	 	           	 	\end{equation} 
          	 	           	 	is a square root process of dimension four with arithmetic mean ${\bf {E}}^{\bar p_t}(Y^*_{\tau^*_t})=1$  
          	 	           	 	that is	evolving in the atom-GOP activity time 
          	 	           	 	\begin{equation}
          	 	           	 		\tau^*_t=\tau^*_0+\int_{0}^{t}a^*_sds
          	 	           	 	\end{equation}
          	 	           	 	and  satisfies the SDE
          	 	           	 	\begin{equation}\label{Y^*_t}
          	 	           	 		d Y^*_{\tau^*_t}=\left(1-Y^*_{\tau^*_t}\right)a^*_tdt+\sqrt{Y^*_{\tau^*_t}a^*_t}dW^*_t
          	 	           	 	\end{equation}
          	 	           	 	with initial value 
          	 	           	 	\begin{equation}
          	 	           	 		Y^*_{\tau^*_0}=S^*_0
          	 	           	 	\end{equation} and  {\em atom-GOP activity} 
          	 	           	 	\begin{equation}\label{a*Zt}
          	 	           	 		a^*_t=a_tZ_tY^*_{\tau^*_t}
          	 	           	 	\end{equation}
          	 	           	 	for $t \in [0,\infty)$.
          	 	           	 \end{theorem}
          	 	           	 The above-described type of model has been suggested in \citeN{Platen97d} as a stochastic volatility model and in \citeN{Platen01a} as the  {\em minimal market model} (MMM); see, e.g.,   \citeN{HulleySc10}. It plays the role of the num\'eraire for benchmark-neutral pricing; see \citeN{Platen25a} and \citeN{SchmutzPlSc25}. When employed for the  pricing and hedging of extreme-maturity pension and insurance contracts, this model  turned out to be a  realistic GOP model, as demonstrated, e.g., in \citeN{FergussonPl23} and \citeN{BaroneadesiPlSa24}.  In \citeN{Platen25a} it has been shown that the hedge error for a zero coupon bond is extremely small, which indicates that the obtained information-minimizing model is a  realistic model that generates hedge errors over several decades that remain extremely small.\\
          	 	           	 
          	 	           	 
          	 	           	 By Theorem \ref{stockGOP1}, the normalized atom GOP $	Y^*_{\tau^*_t}$  is   following in the  atom-GOP activity time $\tau^*_t$ a square root  process of dimension four with mean $1$. This process has as stationary density a gamma density with four degrees of freedom.  As mentioned earlier, the Student-$t$ density with four degrees of freedom is the  normal mixture density that emerges as the log-return density when the mixing stationary density of the squared volatility is that of the inverse of a gamma distributed random variable with four degrees of freedom. If one estimates the log-return density of  the  atom GOP of an information-minimizing market, then Theorem \ref{stockGOP1} predicts the estimation of a Student-$t$ density with four degrees of freedom. In \citeN{PlatenRe08e},  a  proxy of the GOP
          	 	           	 of the investment universe formed by the stocks of the MSCI world stock index was constructed. When this proxy would be interpreted in \citeN{PlatenRe08e} as a proxy of the atom GOP of the world stock market, which is by Theorem \ref{informationtheorem} an equally weighted atom portfolio, then   the hypothesis that the log-returns of the atom GOP have a Student-t  density with four degrees of freedom could not be  rejected  
          	 	           	 on a high  significance level  when tested in a large class of  log-return densities.\\

          	 	           	 \subsection{Minimum Variance Portfolio}
          	 	           	
          	 	           	 In Appendix D the following property of the MVP  is derived:
          	 	           	 \begin{theorem}\label{MVPTheorem1}
          	 	           	 	For an information-minimizing market of atoms, \\
          	 	           	 		the  MVP  equals the AP, that is,
          	 	           	 		\begin{equation}\label{SMVPSAP}
          	 	           	 			S^{MVP}_t=S^{AP}_t=\sum_{k=1}^{n}  A^k_{t},
          	 	           	 		\end{equation}
          	 	           	 		where the squared MVP-volatility satisfies the inequality
          	 	           	 		\begin{equation}\label{MVPineq1}
          	 	           	 			(\sigma^{MVP}_t)^2=	(\sigma^{AP}_t)^2
          	 	           	 			=\frac{ a_t}{Y^{AP}_{\tau^{AP}_t}} 
          	 	           	 			\leq (\sigma^{S^*}_t)^2=Z_ta_t=\frac{a^*_t}{Y^*_{\tau^*_t}}
          	 	           	 		\end{equation}  with $Z_t$ given in \eqref{Zt} and \begin{equation}Y^{AP}_{\tau^{AP}_t}=\frac{S^{AP}_t}{B_te^{\tau^{AP}_t}}\end{equation} given in \eqref{SAP}         	 	           	 		        	 	     
\end{theorem}
The above theorem reveals the fact that the  MVP of an information-minimizing market     equals its AP and forms  an SROU process of dimension four.   This means, its dynamics are  probabilistically of the same type as those of other sums of atoms and those of the atom GOP. For $n>1$, atom GOP and AP dynamics involve distinct activity times and aggregate Brownian motions.   
According to the inequality \eqref{MVPineq1}, the squared volatility, instantaneous growth rate, and expected return of the atom GOP  are  greater than their  counterparts  of the AP. \\ From a portfolio management perspective, the AP reflects a {\em buy-and-hold strategy} where one unit of each atom is retained.    In contrast, the atom GOP is generated via dynamic reallocation, ensuring that each atom is assigned an equal weight.   The dynamical reallocation strategy of the atom GOP, which requires reallocation work, provides a larger instantaneous growth rate and a greater expected return than the buy-and-hold strategy of the AP.  The difference between the average growth rates of both portfolios increases with the number $n$ of atoms. This fact allows one to conclude that it should be possible to achieve higher average growth rates than those of widely popular stock index funds by investing dynamically with adequate constant proportions in the stock market similarly as shown in  \citeN{PlatenRe08e} and \citeN{PlatenRe20}. \\

This prompts the question: what occurs when nearly all atoms are invested in the atom GOP $S^*_t$? According to \eqref{hatlambda} in Theorem \ref{informationtheorem} and Theorem \ref{stockGOP1}, the expected growth rate $G^{S^{AP}}=G^{S^{MVP}}$ of  the AP is expected to closely match the expected growth rate $G^{S^*}$of the atom GOP. 
This is because, under the law of conservation of energy, any additional growth of the atom GOP must have a source, implying that the following relationship holds approximately:
\begin{equation}
G^{S^{MVP}}=(\hat \lambda +1){\bf{E}}(a_t)+G^{A^0}\approx G^{S^*}={\bf{E}}(\lambda^*_t+\frac{a_t}{2}Z_t)= (\hat \lambda+\frac{\bar\omega(n)}{2}){\bf{E}}(a_t)+G^{A^0}.
\end{equation}
Since $\bar\omega(n)>1$ for $n>1$ the following conclusion can be drawn:
\begin{corollary}
	In an information-minimizing market, where almost all atoms are  fully invested in the atom GOP, it follows that the average activity must become almost zero.
\end{corollary}
This means  in the above case that the volatilities are extremely small and the growth rate of the   SMVP and the atom GOP are close to the interest rate.
          	\subsection{Scaling Property}
          	 As pointed out in Chapter XI of  \citeN{RevuzYo99}, squared Bessel processes have a  {\em scaling property}, which is  a self-similarity property  in the sense of \citeN{Mandelbrot97}.  This scaling property  arises from a Lie-group symmetry of the transition probability density of a squared Bessel process. It  conserves the type of probability density concerning value and  time.  By application of the It\^o formula, it follows that atoms, sums of atoms, the MVP,  the atom GOP, and the GOP of the extended market when denominated in units of the basis exponential $B_t$ (given in \eqref{basis}) or in the case of the extended market GOP denominated by the savings account, represent time-transformed squared Bessel processes. Therefore, all these portfolios have the following scaling property:
          		\begin{corollary}\label{scaling}
          			For  an information-minimizing market, every basis exponential-denominated atom portfolio $\bar A^{\mathcal{A}}_{\varphi^{\mathcal{A}}(t)}=\frac{A^{\mathcal{A}}_t}{B_t}$ with   $\mathcal{A}\subseteq \{1,...,n\}$, dimension $d_{\mathcal{A}}$, and intrinsic time $\varphi^{\mathcal{A}}(t)$ has the {\em scaling property} that for every {\em scaling} $c>0$, the process $c^{-1}\bar A^{\mathcal{A}}_{c\varphi^{\mathcal{A}}(t)}$ represents a time-transformed squared Bessel process with the original dimension $d_{\mathcal{A}}$, scaled initial value $c^{-1}\bar A^{\mathcal{A}}_{c\varphi^{{\mathcal{A}}}(0)}$, and scaled  intrinsic time $c\varphi^{{\mathcal{A}}}({t})$ for $t\in[0,\infty)$. Similarly, the basis exponential-denominated atom GOP $\bar S^*_{\varphi^*(t)}=\frac{S^*_t}{B_t}$ and the savings account-denominated extended market GOP  $\bar S^{**}_{\varphi^{**}(t)}=\frac{S^{**}_t}{A^0_t}$  have an analogous scaling property.
          		\end{corollary}
          	The scaling properties for $\bar S^{*}_{\varphi^{*}(t)}$ and $\bar S^{**}_{\varphi^{**}(t)}$	follow from the fact that these processes are squared Bessel processes in respective intrinsic times. The scaling property described above accounts for the self-similarity in stock indices noted by \citeN{Mandelbrot97}.   This property  is  consistent with the scaling property of log-returns of stock indices that was empirically identified in \citeN{BreymannLuPl09e}. 


         \section*{Conclusion}
         Based on  two mathematical principles, 
          the paper derived a realistic, parsimonious model for the natural, undisturbed  dynamics of  the financial market by minimizing the  information of the stationary density of the normalized atoms and the average squared market prices of risk. The findings indicate that financial markets can be viewed as communication systems, and that concepts from information theory are particularly pertinent to the field of finance.  The derived results  represent  the first steps in a promising research direction,  which will open new avenues for resolving challenging problems in finance and economics. Forthcoming work will provide more results and empirical evidence in this direction. 

       \appendix
        
        \textwidth14.5cm \textheight9in
        \topmargin0pt
        \renewcommand{\theequation}{\mbox{B.\arabic{equation}}}
        \renewcommand{\thefigure}{B.\arabic{figure}}
        \renewcommand{\thetable}{B.\arabic{table}}
        \renewcommand{\footnoterule}{\rule{14.8cm}{0.3mm}\vspace{+1.0mm}}
        \renewcommand{\baselinestretch}{1.0}
        \pagestyle{plain}

        \textwidth14.5cm \textheight9in
        \topmargin0pt
        \renewcommand{\theequation}{\mbox{A.\arabic{equation}}}
        \renewcommand{\thefigure}{A.\arabic{figure}}
        \renewcommand{\thetable}{A.\arabic{table}}
        \renewcommand{\footnoterule}{\rule{14.8cm}{0.3mm}\vspace{+1.0mm}}
        \renewcommand{\baselinestretch}{1.0}
        \pagestyle{plain}

      \section*{Appendix A: Proof of Theorem \ref{stockGOP}}\setcA \setcB 
      By applying  Theorem 3.1 in \citeN{FilipovicPl09} for $t\in[0,\infty)$, we obtain from  \eqref{hatB11j21} with the SDE (3.10)   in \citeN{FilipovicPl09}  
       for the portfolio $S^{\bar\pi}_t$ with weight vector $\bar\pi_t=(\bar\pi^1_t,...,\bar\pi^n_t)^\top$ for the holdings in the atoms ${\bf{A}}_t$ the SDE
      	\BE \label{e.4.12}
      	\frac{dS^{\bar\pi}_t}{S^{\bar\pi}_t}=(\bar\pi_t)^\top\frac{d{\bf{A}}_t}{{\bf{A}}_t}= \lambda^*_t dt +
      	\sum_{k=1}^{n}\bar\pi^{k}_t\beta^k_t(\beta^k_t \omega^k dt+d{ W}^k_t). \EE
       	By the matrix equation (3.5)   in \citeN{FilipovicPl09}  it follows  for the $k$-th atom GOP weight
       	\begin{equation}
       	(\beta^k_t)^2\bar\pi^{*,k}_t+\lambda^*_t= \lambda^*_t+	(\beta^k_t)^2\omega^k,
       	\end{equation}
       	which yields
       	\begin{equation}\label{pik}
       	\bar\pi^{*,k}_t=\omega^k
       	\end{equation}
    for $k\in\{1,...,n\}$ and $t\in[0,\infty)$.   	By 
     applying Equation (3.8) in   \citeN{FilipovicPl09} we obtain the $k$-th atom GOP volatility
       	\begin{equation}\label{theta4}
       	 \bar\pi^{*,k}_t\beta^k_t=\omega^k\beta^k_t
       	\end{equation}
       	for  $k\in\{1,...,n\}$ and $t\in[0,\infty)$. Furthermore, we obtain by Equation (3.4)   in \citeN{FilipovicPl09} and \eqref{pik}
       	\begin{equation}
       	\sum_{k=1}^{n}\omega^k=\sum_{k=1}^{n}\bar\pi^{*,k}_t=1,
       	\end{equation}
        which proves Equation \eqref{Nt=1}. By \eqref{pik} one obtains the Equation \eqref{pi1} and the SDE \eqref{e.4.122a}, which completes  the proof of Theorem \ref{stockGOP}.\\
       	
       	\textwidth14.5cm \textheight9in
       	\topmargin0pt
       	\renewcommand{\theequation}{\mbox{B.\arabic{equation}}}
       	\renewcommand{\thefigure}{B.\arabic{figure}}
       	\renewcommand{\thetable}{B.\arabic{table}}
       	\renewcommand{\footnoterule}{\rule{14.8cm}{0.3mm}\vspace{+1.0mm}}
       	\renewcommand{\baselinestretch}{1.0}
       	\pagestyle{plain}
       	 \section*{Appendix B: Proof of Theorem \ref{GOPentire}}\setcA \setcB
       	
       	Since the savings account $A^0_t$ is a traded security in the extended market, it follows from Theorem 3.1 in \citeN{FilipovicPl09} that the  risk-adjusted return of this market is the interest rate $r_t$. For a currency-denominated portfolio $S^{\pi}_t$, which invests with the weight $ \pi^{0}_t$ in the savings account $A^0_t$ and with the weight $ \pi^k_t$ in the $k$-th atom $A^k_t$, $k\in\{1,...,n\}$, it follows by Equation (3.11) in \citeN{FilipovicPl09} that this portfolio equals the extended-market GOP $S^{**}_t$  when it satisfies the SDE
       	\begin{equation}
       		\frac{dS^{**}_t}{S^{**}_t}=r_tdt+\sum_{k=1}^{n}\theta^{k}_t\left(\theta^{k}_tdt+dW^k_t\right)
       	\end{equation}
       	with  \begin{equation}
       		\theta^{k}_t=\pi^{**,k}_t\beta^k_t,
       	\end{equation}
       	which proves  equation  \eqref{bare.4.113}. By Equation (3.8) in  \citeN{FilipovicPl09} we have the  $k$-th market price of risk
       	 and by the  Equation (3.5) in  \citeN{FilipovicPl09}  the equation
       	\begin{equation}
       		(\beta^k_t)^2\pi^{**,k}_t= \lambda^*_t-r_t+	(\beta^k_t)^2\omega^k,
       	\end{equation} which is solved by the optimal $k$-th weight
       	\begin{equation}\label{barpi**}
       	\pi^{**,k}_t=\frac{ \lambda^*_t-r_t}{	(\beta^k_t)^2}+\omega^k
       	\end{equation}
       	and yields the $k$-th market price of risk
       	\begin{equation}
       		\theta^{k}_t=\frac{ \lambda^*_t-r_t}{	\beta^k_t}+\omega^k\beta^k_t
       	\end{equation}
       	for $k\in\{1,...,n\}$ and $t\in[0,\infty)$.	 We obtain by \eqref{barpi**} and \eqref{Nt=1} 
       	the weight
       	\begin{equation}
       		\pi^{**,0}_t=1-\sum_{k=1}^{n}\pi^{**,k}_t=1- (\lambda^*_t-r_t)\sum_{k=1}^{n}(\beta^k_t)^{-2}-1= - (\lambda^{*}_t-r_t)\sum_{k=1}^{n}(\beta^k_t)^{-2}
       	\end{equation}
       	to be invested  in the savings account $A^0_t$. 
       	This completes the proof of Theorem \ref{GOPentire}.

       	\textwidth14.5cm \textheight9in
       	\topmargin0pt
       	\renewcommand{\theequation}{\mbox{C.\arabic{equation}}}
       	\renewcommand{\thefigure}{C.\arabic{figure}}
       	\renewcommand{\thetable}{C.\arabic{table}}
       	\renewcommand{\footnoterule}{\rule{14.8cm}{0.3mm}\vspace{+1.0mm}}
       	\renewcommand{\baselinestretch}{1.0}
       	\pagestyle{plain}
       	\section*{Appendix C: Proof of Theorem \ref{informationtheorem}}\setcA \setcB
       	We perform the minimization of the joint information
       	\begin{equation}
       		\mathcal{I}(p_{t},	\Lambda_t)=	\mathcal{I}(p_{0},	p_{0})+\int_{0}^{t}I(p_{s},	q_{s})ds,
       	\end{equation} in six steps for $t\in[0,\infty)$.\\
       	
       1. First we derive the stationary densities of the normalized atoms.	For $k\in\{1,...,n\}$, the stationary density $p^k_0=p^k_t=p^k_\infty$   of the $k$-th normalized atom, which is evolving in the $k$-th  activity time $\tau^k$,
       	is by the SDE \eqref{bardYtau} the  solution of the  stationary Fokker-Planck equation 
       	\begin{equation}
       \frac{d}{dy}	 \left(p^k_\infty(y)y((\phi^k(y))^{-1}\omega^k-1)\right) -\frac{1}{2}\frac{d^2}{dy^2} \left(p^k_\infty(y)y^2\frac{1}{\phi^k(y)}\right) =0,
       	\end{equation} as described, e.g.,in  Chapter 4 in \citeN{PlatenHe06}, which is a  second-order ordinary differential equation. Its solution is given by the formula 
       	\BE \label{qjy}
       	p^{k}_\infty(y)=\frac{C_k\phi^k(y)}{y^2 }\exp \left\{2\int_{1}^{y}\frac{\omega^k-\phi^k(u)}{u}du\right\}
       	\EE
       	for $ y\in(0,\infty)$ and some constant $C_k>0$. The latter ensures that $p^k_\infty$ is a probability density. \\
       	
       2.	Under the  constraints \eqref{barY0*} and \eqref{barstalogmean*}   we minimize  the sum  \begin{equation}\mathcal{I}(p_\infty,p_\infty)=\sum_{k=1}^{n}{\mathcal{I}}(p^{k}_\infty ,p^{k}_\infty)\end{equation}  of the information of the stationary probability densities $p^{k}_\infty$ of the independent normalized atoms $Y^k_{\tau^k_.} $, $k\in\{1,...,n\}$. According to the formula \eqref{Ipq} we minimize the  Lagrangian
       	\begin{equation*}
       	{\cal{L}}(p^{k}_\infty, \lambda_0,\lambda_1,\lambda_2)= 
       	\int_{0}^{\infty} p^k_\infty (y)\ln(p^k_\infty (y))dy-\lambda_0 \left(\int_{0}^{\infty}p^{k}_\infty(y)dy-1\right)\end{equation*}\BE -\lambda_1\left(\int_{0}^{\infty}y p^{k}_\infty(y)dy-\omega^k\right)
       	-\lambda_2\left(\int_{0}^{\infty}\ln(y)p^{k}_\infty(y)dy-\zeta^k\right),
       	\EE
       	where $\lambda_0, \lambda_1, \lambda_2$ are Lagrange multipliers. $ {\cal{L}}(p^{k}_\infty, \lambda_0,\lambda_1,\lambda_2)$ is minimized when its Fr\'echet derivative $\delta {\cal{L}}(p^{k}_\infty, \lambda_0,\lambda_1,\lambda_2)$, i.e., the first variation of ${\cal{L}}(p^{k}_\infty, \lambda_0,\lambda_1,\lambda_2)$ with respect to admissible variations of $p^{k}_\infty$, becomes zero. This implies for the information-minimizing stationary density $\bar p^{k}_\infty$ the equation
       	\BE
       	\delta {\cal{L}}(\bar p^{k}_\infty, \lambda_0,\lambda_1,\lambda_2)=\int_{0}^{\infty}\left(\ln(\bar p^{k}_\infty(y))-\lambda_0-\lambda_1 y-\lambda_2 \ln(y)\right)\delta \bar p^{k}_\infty(y) dy=0.
       	\EE
       	The solution of the above first-order condition is the gamma density 
       	\BE \label{barpy}
       	\bar p^{k}_\infty(y)=\exp \{\lambda_0+ \lambda_1 y+\lambda_2 \ln(y)\}
       	\EE
       	for $y\in (0,\infty)$ with the constraint \begin{equation}
       	\int_{0}^{\infty}\exp \{\lambda_0+ \lambda_1 y+\lambda_2 \ln(y)\}dy=1,
       	\end{equation}
       	and the Lagrange multipliers $\lambda_0, \lambda_1,\lambda_2$. It
       	has  $2(\lambda_2+1)$ degrees of freedom and it  parametrizes the averages \BE {\bf{E}}^{\bar p^{k}_\infty}(Y^k_{.})=\frac{\lambda_2+1}{-\lambda_1}=\omega^k\EE 
       	and
       	$$ {\bf{E}}^{\bar p^{k}_\infty}(\ln(Y^k_{.}))=\zeta^k.$$
       	\\On the other hand, the SDE for the   $k$-th normalized atom  is given  by \eqref{bardYtau}.  
       	Consequently, the stationary density $p^{k}_\infty(y)$ of the $k$-th normalized atom satisfies the Fokker-Planck equation  with the drift and diffusion coefficient functions of the SDE \eqref{bardYtau}. This yields  the stationary density $p^{k}_\infty(y)$ in the form  \eqref{qjy}. The latter must equal the above-identified gamma density. By setting  both expressions for the stationary density equal, respective conditions for the  function $\phi^k(y)$ emerge.\\
       	The Weierstrass Approximation Theorem states that a continuous function can be approximated on a bounded interval by a polynomial.  When using a polynomial for characterizing $\phi^k(y)$ and searching for a match of the stationary density \eqref{qjy} with the gamma density  \eqref{barpy},  one finds by comparing the coefficients of the possible polynomials  that only the polynomial
       	\BE\label{psi1}
       	\phi^k(y)=y 
       	\EE
       	provides such a match. 
      	This yields for the $k$-th normalized atom process $Y^k_.$ the 
       	stationary density
       	\BE\label{qY1}
       	p^{k}_\infty(y)=\frac{C_k y }{y^2 }\exp \left\{2\int_{1}^{y}\frac{\omega^k-u }{u}du\right\}=\frac{2^{2\omega^k}y^{2\omega^k-1}}{\Gamma(2\omega^k)} \exp\{-2y\}
       	\EE
       	for $y>0$. The above density  is the gamma density with $d_k=4\omega^k$ degrees of freedom and mean $\omega^k$. We assumed
       	the logarithmic average of the stationary density to equal   a constant $\zeta^k$, which emerges as
       	\BE\label{Hq'}
       	\zeta^k=  {\bf{E}}^{ p^{k}_\infty}(\ln(Y^k_{.}))=\ln\left(\frac{1}{2}\right)+  \psi(2\omega^k),
       	\EE
       	where the function $\psi(x)$ 
       	is the diagamma function
       	. 
       	The resulting square root process is a stationary process when its initial value $Y^k_{0}=A^k_0$ is distributed according to its stationary density. The self-information of the $k$-th stationary density equals
       	\begin{equation}
       		\mathcal{I}(p^k_\infty,p^k_\infty) = \int_{0}^{\infty}\ln(p^k_\infty(y))p^k_\infty(y)dy\end{equation}\begin{equation}=\ln\left(\frac{2^{2\omega^k}}{\Gamma(2\omega^k)}\right)+(2\omega^k-1){\bf{E}}^{ p^{k}_\infty}(\ln(Y^k_{.}))-2{\bf{E}}^{ p^{k}_\infty}(Y^k_{.})
       	\end{equation}
       	\begin{equation}=(2\omega^k-1)(  \psi(2\omega^k)-\ln(2))-2\omega^k-\ln\left(\Gamma(2\omega^k)\right)+2\omega^k\ln(2),
       	\end{equation}
       	which yields
       	\begin{equation}\mathcal{I}(p_\infty,p_\infty)=\sum_{k=1}^{n}\left((2\omega^k-1)  \psi(2\omega^k)+\ln(2)-2\omega^k-\ln\left(\Gamma(2\omega^k)\right)\right).
       	\end{equation}
       3. In the next step we minimize the Kullback-Leibler divergence  of $q_t$ with respect to $p^{}_t$: 
              \begin{equation}I( p_{t},q^{}_t)=-\sum_{k=1}^{n}\frac{d}{dt}{\bf{E}}\left(\ln(\Lambda^{}_t)\right)
       	=\frac{1}{2}\sum_{k=1}^{n}{\bf{E}}\left((\theta^k_t)^2\right)
       	=G^{\frac{S^{**}}{A^0}} \rightarrow \min,\end{equation}	
     which  equals by Equation \eqref{thetak}  the	 expected growth rate $G^{\frac{S^{**}}{A^0}}$ of the extended market GOP when denominated in the savings account.
      This expected	 growth rate equals the sum
       \begin{equation}G^{\frac{S^{**}}{A^0}}=G^{\frac{S^{**}}{S^*}}+G^{\frac{S^{*}}{B}}+G^{\frac{B}{A^0}}\end{equation}		
       of the expected growth rate $G^{\frac{S^{**}}{S^*}}$ of the extended market GOP denominated in the atom GOP, the expected growth rate  $G^{\frac{S^{*}}{B}}$ of the atom GOP  denominated in the basis exponential, and the expected growth rate $G^{\frac{B}{A^0}}$ of the basis exponential when denominated in the savings account. In the following we minimize step by step each of these three expected growth rates.\\
       
       4. In \eqref{averactivity}  the average activity is defined as 
       \begin{equation*}
       	a_t=
       	\left(\sum_{k=1}^{n}\omega^k\sqrt{\frac{1}{a^k_t}}\right)^{-2}.
       \end{equation*}
       This allows us to write the expected growth rate $G^{\frac{S^{**}}{S^*}}$ in the form
       \begin{equation*}
       	G^{\frac{S^{**}}{S^*}}=
       {\bf{E}}\left(	\frac{(\lambda^*_t- r_t)^2}{2}\sum_{k=1}^{n}\frac{{\bf{E}}^{ p^k_{t}}\left(Y^k_{\tau^k_t}\right)}{a^k_t}\right)= {\bf{E}}\left(	\frac{(\lambda^*_t- r_t)^2}{2}\sum_{k=1}^{n}\frac{\omega^k}{a^k_t}\right)
       \end{equation*} 
       \begin{equation*}
       	= {\bf{E}}\left(\frac{(\lambda^*_t- r_t)^2}{2}\left(\frac{1}{ a_t}+\sum_{k=1}^{n}\omega^k\left(\sqrt{\frac{1}{a^k_t}}-\sqrt{\frac{1}{ a_t}}\right)^2 \right)\right).
       \end{equation*}
       This expected growth rate is minimized with respect to the choice of the activities if all activities are equal, which proves \eqref{akt} and yields
       \begin{equation}\label{GSS1}
       	G^{\frac{S^{**}}{S^*}}
       	= {\bf{E}}\left(\frac{(\lambda^*_t- r_t)^2}{2a_t}\right).
       \end{equation}
        \\
       5.  We introduce the average risk premium factor
       \begin{equation*}
       	\bar\omega(n)=\left(\sum_{k=1}^{n}\omega^k\sqrt{{\bf{E}}^{ p^k_{t}}\left(\frac{ \omega^k }{Y^k_{\tau^k_t}}\right)}\right)^2.
       \end{equation*}
       The expected growth rate of the atom GOP $S^*_t$ when denominated in the basis exponential $B_t$ equals
       \begin{equation*}
       	G^{\frac{S^{*}}{B}}={\bf{E}}\left(\frac{ a_t}{2}Z_t\right)={\bf{E}}\left(\frac{ a_t}{2}\sum_{k=1}^{n}\omega^k{\bf{E}}^{ p^k_{t}}\left(\frac{ \omega^k }{Y^k_{\tau^k_t}}\right)\right)
       \end{equation*}
       \begin{equation*}
       	={\bf{E}}\left(\frac{ a_t}{2}\sum_{k=1}^{n}\omega^k\left(\sqrt{\bar\omega(n)}+\left(\sqrt{{\bf{E}}^{ p^k_{t}}\left(\frac{ \omega^k }{Y^k_{\tau^k_t}}\right)}-\sqrt{\bar\omega(n)}\right)\right)^2\right)
       \end{equation*}
       \begin{equation*}
       	={\bf{E}}\left(\frac{ a_t}{2}\left(\bar\omega(n)+\sum_{k=1}^{n}\omega^k\left(\sqrt{{\bf{E}}^{ p^k_{t}}\left(\frac{ \omega^k }{Y^k_{\tau^k_t}}\right)}-\sqrt{\bar\omega(n)}\right)^2\right)\right) 
       \end{equation*}
       This expected growth rate is minimized when all $\omega^k$ are equal, which means by \eqref{Nt=1} that we have
       \begin{equation}
       	\omega^k=\frac{1}{n}
       \end{equation}
       and 
       \begin{equation}
       	\bar\omega(n)={\bf{E}}^{ p^k_{t}}\left(\frac{\frac{1}{n}}{Y^k_{\tau^k_t}}\right)={\bf{E}}^{ p_{t}}\left(Z_t\right)
       \end{equation}
       for $k\in\{1,...,n\}$. This proves \eqref{omega} and \eqref{baromega}. Furthermore, we get
       \begin{equation}\label{GS*B1}
       	G^{\frac{S^{*}}{B}}={\bf{E}}\left(\frac{\bar \omega(n)  a_t}{2}\right), 
       \end{equation}
             and have by \eqref{A0} and \eqref{basis} the expected growth rate
       \begin{equation}\label{GBA01}
       	G^{\frac{B}{A^0}}={\bf{E}}\left(	\lambda^*_t-r_t\right).	\end{equation} \\
       6. By summing up \eqref{GSS1}, \eqref{GS*B1}, and \eqref{GBA01}, we obtain the Kullback-Leibler divergence in the form
       	\begin{equation*}I( p_{t},q^{}_t)
       		={\bf{E}}\left(\frac{ a_t}{2}\left(\frac{\left(\lambda^*_t-  r_t\right)^2}{(a_t)^2}+\bar \omega(n) +\frac{2(\lambda^*_t- r_t)}{a_t}+{\bf{E}}^{p_t}\left(\left(\frac{\lambda^*_t- r_t}{a_t}-\hat\lambda\right)^2\right)\right)\right).\end{equation*}
      The Kullback-Leibler divergence becomes fully minimized when setting \begin{equation}\hat \lambda=\frac{\lambda^*_t-r_t}{a_t},\end{equation}
      which yields 
       \begin{equation*}I( p_{t},q^{}_t)={\bf{E}}\left(\frac{ a_t}{2} \left(\hat \lambda^2+\bar \omega(n) +2\hat \lambda\right)\right)\end{equation*}
       for $t\in[0,\infty)$. This  proves \eqref{relinfo} and, therefore, Theorem \ref{informationtheorem}. 
       	 \textwidth14.5cm \textheight9in
       	 \topmargin0pt
       	 \renewcommand{\theequation}{\mbox{D.\arabic{equation}}}
       	 \renewcommand{\thefigure}{D.\arabic{figure}}
       	 \renewcommand{\thetable}{D.\arabic{table}}
       	 \renewcommand{\footnoterule}{\rule{14.8cm}{0.3mm}\vspace{+1.0mm}}
       	 \renewcommand{\baselinestretch}{1.0}
       	 \pagestyle{plain}
       	 \section*{Appendix D: Proof of Theorem \ref{stockGOP1}}\setcA \setcB

       	       		For $t\in[0,\infty)$, by employing Theorem \ref{stockGOP} with \eqref{e.4.122a}, \eqref{beta}, 
       	       	and	\eqref{psi1}  
       	       		   it follows   the SDE
       	       		\begin{equation} \label{e.4.1222}
       		\frac{dS^{*}_t}{S^{*}_t}
       		= \lambda^*_t dt +
       		\sum_{k=1}^{n}\sqrt{\frac{(\omega^k)^2 a^k _t}{Y^k_{\tau^k_t}}}\left(\sqrt{\frac{(\omega^k)^2 a^k _t}{Y^k_{\tau^k_t}}} dt+d{ W}^k_t\right). \end{equation}
        For  the  atom GOP $S^*_t$ we can rewrite its SDE with \eqref{akt} in the form
       	         	\begin{equation} \label{e.4.119}
       	d  S^*_{t}= \lambda^*_t S^*_{t}dt+
       		\sum_{k=1}^{n}\frac{ S^*_{t}(\omega^k)^2 a _t}{Y^k_{\tau^k_t}}dt
       	+	\sum_{k=1}^{n}	\sqrt{ S^*_{t}}\sqrt{\frac{ S^*_{t}(\omega^k)^2 a_t}{Y^k_{\tau^k_t}}} d{ W}^k_t.
       \end{equation}
       With \eqref{Zt} we have 
       \begin{equation}\label{Zt1}
       Z_t=  \sum_{k=1}^{n}\frac{(\omega^k)^2 }{Y^k_{\tau^k_t}},
       \end{equation} and with
       \eqref{W*t}  the Brownian motion $W^*_t$ with stochastic differential
       \begin{equation}
       dW^*_t= Z_t^{-\frac{1}{2}}\sum_{k=1}^{n}\sqrt{\frac{ (\omega^k)^2}{Y^k_{\tau^k_t}}}d W^k_t
       \end{equation} with initial value $W^*_0=0$. This allows us to       
       introduce the  derivative
        \begin{equation}\label{dvarphi}
       \frac{d{\varphi^*(t)}}{dt}=	\sum_{k=1}^{n}\frac{ S^*_{t}(\omega^k)^2 a_t}{4Y^k_{\tau^k_t}}= \frac{a_t}{4} S^*_{t}Z_t
       \end{equation}
     of the intrinsic atom GOP time $\varphi^*(t)$, and to rewrite the SDE \eqref{e.4.119} in the form
       		\BE \label{e.4.120}
       		d  S^*_{t}
       		 = \lambda^*_t S^*_{t}dt+
       		 4	\frac{d{\varphi^*(t)}}{dt}dt+ \sqrt{ S^*_{t}}  \sqrt{4\frac{d{\varphi^*(t)}}{dt}} dW^*_t,
       		\EE
       		which proves the SDE \eqref{e.4.118} together with \eqref{Zt}, 
       		 \eqref{W*t}, and \eqref{varphi*}.\\
       		 
        Similarly as in \eqref{Yktayt1} we can introduce the normalized atom GOP
           \begin{equation}\label{Y*}
           Y^*_{\tau^*_t}=\frac{ S^*_{t}}{B_te^{\tau^*_t-\tau^*_0}}
           \end{equation}  with $Y^*_{\tau^*_t}$ 
           forming a square root process of dimension four with arithmetic average ${\bf {E}}^{\bar p_t}(Y^*_{\tau^*_t})=1$  that
           evolves in the atom-GOP activity time 
         \begin{equation}
          \tau^*_t=\tau^*_0+\int_{0}^{t}a^*_sds
         \end{equation}
       according to the SDE
         \begin{equation}
         	\frac{d Y^*_{\tau^*_t}}{Y^*_{\tau^*_t}}=\left(\frac{Z_t}{a^*_t}-1\right)a^*_tdt+\sqrt{Z_t}dW^*_t=\left(\frac{1}{Y^*_{\tau^*_t}}-1\right)a^*_tdt+\sqrt{\frac{a^*_t}{Y^*_{\tau^*_t}}}dW^*_t
         \end{equation}
          with the {\em atom-GOP activity} 
                    \begin{equation}
         a^*_t=Z_tY^*_{\tau^*_t}
         \end{equation} 
       	for $t\in[0,\infty)$. This proves the remaining statements of Theorem \ref{stockGOP1}.

          		 \textwidth14.5cm \textheight9in
          		\topmargin0pt
          		\renewcommand{\theequation}{\mbox{E.\arabic{equation}}}
          		\renewcommand{\thefigure}{E.\arabic{figure}}
          		\renewcommand{\thetable}{E.\arabic{table}}
          		\renewcommand{\footnoterule}{\rule{14.8cm}{0.3mm}\vspace{+1.0mm}}
          		\renewcommand{\baselinestretch}{1.0}
          		\pagestyle{plain}
          		\section*{Appendix E: Proof of Theorem \ref{MVPTheorem1}}\setcA \setcB
          		We partition the proof into three parts:\\
          		
          		1. For $t\in[0,\infty)$ and $k\in\{1,...,n\}$, 
          		we have by Equation \eqref{hatB11j21} for the  $k$-th  atom $
          		A^k_{t}$
          		the SDE 
          		\begin{equation}\label{barAkSDE}
          			\frac{d   A^k_t}{ A^k_t}
          			=\lambda^*_tdt+\beta^k_t\left(\beta^k_t\omega^k_tdt+dW^k_t\right) 
          		\end{equation}
          		with initial value
          		$A^k_0>0$.
          		A portfolio $S^{\pi}_{t}$ of  atoms with weight vector $ \pi_t=(\pi^1_t,...,\pi^{n}_t)^\top$ invested in $A^1_{t},..., A^n_{t}$ satisfies by \eqref{barAkSDE}  the SDE
          		\begin{equation}\label{portfolioSDE1}
          		\frac{d S^{\pi}_{t}}{ S^{\pi}_{t}}=\sum_{k=1}^{n}\pi^k_t \frac{d   A^k_{t}}{  A^k_{t}} 
          		=\lambda^*_tdt+\sum_{k=1}^{n}\pi^k_t\beta^k_t\left(\beta^k_t\omega^k_tdt+dW^k_t\right) .
          		\end{equation}
          		The squared volatility of the  portfolio $ S^{\pi}_{t}$ equals
          		\begin{equation}\label{sigmaMVP}
          		( \sigma^{\pi}_t)^2=\sum_{k=1}^{n}\left(\pi^k_t\beta^k_t\right)^2  
          		.
          		\end{equation} 
          		According to Definition \ref{MVP}, to identify the MVP, we minimize $ ( \sigma^{\pi}_t)^2$  and employ the  Lagrangian
          		\begin{equation}
          		\mathcal{L}(\pi_t,\lambda_t)=\sum_{k=1}^{n}\left(\pi^k_t\beta^k_t\right)^2  -\lambda_t\left(\sum_{l=1}^{n}\pi^l_t-1\right)
          		\end{equation}
          		with Lagrange multiplier $\lambda_t$.	We obtain  the first order condition
          		\begin{equation}
          		\frac{\partial\mathcal{L} (\pi_t,\lambda_t)}{\partial \pi^k_t}=2\pi^{MVP,k}_t  (\beta^k_t)^2-\lambda_t=0,
          		\end{equation}
          		which yields the {\em $k$-th MVP-weight}
          		\begin{equation}\label{piMVP}
          	\pi^{MVP,k}_t=\frac{\lambda_t}{2 (\beta^k_t)^{2} }
          		\end{equation}
          		for $k\in\{1,...,n\}$. 
          		By \eqref{piMVP} 
          		 and the constraint $\sum_{k=1}^{n}\pi^{MVP,k}_t=1$ we have
          		\begin{equation}
          		1=     \sum_{k=1}^{n}\pi^{MVP,k}_t=\frac{\lambda_t}{2}\sum_{k=1}^{n}(\beta^k_t)^{-2}, 
          		\end{equation}
          which is	yielding 
          		the  Lagrange multiplier
          		\begin{equation}\label{lambda_t}
          		\lambda_t=   \frac{  2}{\sum_{k=1}^{n}(\beta^k_t)^{-2} }.
          		\end{equation}
          		This provides by  \eqref{piMVP} 
          		and \eqref{lambda_t} for $k\in\{1,...,n\}$ the $k$-th MVP weight
          		\begin{equation}\label{piMPk}
          		\pi^{MVP,k}_t= \frac{(\beta^k_t)^{-2}}{\sum_{l=1}^{n}(\beta^l_t)^{-2} }. 
          	\end{equation}\\
          	
        2. It follows by \eqref{beta}, \eqref{akt}, \eqref{psi1},  \eqref{YAP},  \eqref{SAP}, and \eqref{Akt2} the  $k$-th MVP weight in the form
        	\begin{equation}\label{piMPk1}
        	\pi^{MVP,k}_t= \frac{\frac{Y^k_{\tau^k_t}}{a_t}}{\sum_{l=1}^{n}\frac{Y^l_{\tau^l_t}}{a_t} }= \frac{Y^k_{\tau^k_t}}{\sum_{l=1}^{n}Y^l_{\tau^l_t} }= \frac{Y^k_{\tau^k_t}B_te^{\int_{0}^{t}a_sds}}{S^{AP}_{t} }=\frac{A^k_{t}}{S^{AP}_{t} }=\pi^{AP,k}_t. 
        	\end{equation}
        Since we have the same weights for $S^{AP}_t$ and $S^{MVP}_t$ and by Definition \ref{MVP} the same initial values  $S^{MVP}_0=S^{AP}_0$ for both portfolios, it follows 
        \begin{equation}
        	S^{MVP}_t=S^{AP}_t
        \end{equation}
       for all $t\in[0,\infty)$,  which proves \eqref{SMVPSAP}.\\

3.  Since the MVP has the minimal possible squared volatility, 
it follows by 
 \eqref{Ykappa5}, \eqref{Y^*_t}, and \eqref{a*Zt} the inequality
\begin{equation}\label{MVPineq}
(\sigma^{MVP}_t)^2= (\sigma^{AP}_t)^2=\frac{ a_t}{Y^{AP}_{\tau^{AP}_t}}\leq (\sigma^{S^*}_t)^2=\frac{ a^*_t}{Y^{*}_{\tau^*_t}}=Z_ta_t
\end{equation} 
for $t\in[0,\infty)$, which  proves the remaining statements of Theorem \ref{MVPTheorem1}.\\

\bibliographystyle{chicago}
\bibliography{my}          
      
\newpage

 \end{document}